\documentstyle[epsfig]{mn}
\begin{document}

\title
[Completely dark galaxies: their existence, properties, and strategies for
finding them]
{Completely dark galaxies: their existence, properties, and strategies for
finding them}
\author[Neil Trentham, 
Ole M\"oller and Enrico Ramirez-Ruiz] 
{
Neil Trentham,$^{1}$
Ole M\"oller$^{2}$ and Enrico Ramirez-Ruiz$^{1}$ \\ 
$^1$ Institute of Astronomy, Madingley Road, Cambridge, CB3 0HA.\\
$^2$ Cavendish Laboratory, Madingley Road, Cambridge, CB3 0HE.
}
\maketitle 

\begin{abstract} 
{There are a number of theoretical and observational hints 
that large numbers of low-mass
galaxies composed entirely of dark
matter exist in the field.  The theoretical considerations follow
from the prediction of cold dark matter theory that there exist
many low-mass galaxies for every massive one. 
The observational considerations follow 
from the observed paucity of these 
low-mass galaxies in the field but not in dense clusters of galaxies; this
suggests that the lack of small galaxies in the field is due to the
inhibition of star formation in the galaxies as opposed to the fact
that their small dark matter halos do not exist.  In this work we 
outline the likely properties of low-mass dark galaxies, and
describe observational strategies for finding them, and where in
the sky to search.  The results are presented as a function of the global
properties of dark matter, in particular the presence or absence of a
substantial baryonic dark matter component.  If the dark matter is
purely cold and has a Navarro, Frenk \& White density profile, directly
detecting dark galaxies will only be feasible with
present technology if 
the galaxy has a maximum velocity dispersion in excess of 70
km s$^{-1}$, in which case the dark galaxies
could strongly lens background objects.  This is much higher than the
maximum
velocity dispersions in most dwarf galaxies.  If the dark matter in galaxy
halos has a baryonic component close to the cosmic ratio, the possibility
of 
directly detecting dark galaxies is much more realistic; the optimal method of
detection will depend on the nature of the dark matter. 
A number of more indirect methods are also discussed.}
\end{abstract} 

\begin{keywords}  
dark matter -- 
cosmology: observations --
cosmology: theory 
\end{keywords} 

\section{Introduction: evidence for the existence of dark galaxies} 
 
It has been a long-recognized feature of hierarchical clustering models
of galaxy formation, like the cold dark matter (CDM) model, that 
they predict the existence of a very large
numbers of low-mass galaxies.  This follows
directly from the shape of the fluctuation spectrum.  On the mass scales
that correspond to those of galaxies,
cold dark matter
has a fluctuation spectrum $\delta (k)$ whose power spectrum
$\vert \delta (k) \vert^2 \sim k^n$ has index $n
\sim -2$ (Bardeen et al.~1986);   
such a power spectrum results in a mass function of dark-matter
(DM) halos 
that is steep at low masses, close to  
$n(M) \sim M^{-2}$ 
(Press \& Schechter 1974, White \& Rees 1978, White \& Frenk 1991, 
Lee \& Shandarin 1999).
The prediction of CDM theory that there should exist many satellite
galaxies per giant galaxy has recently been confirmed by numerical
simulations (Moore et al.~1999, Klypin et al.~1999 and references therein). 

But observations (Cowie et al.~1996,
Lin et al.~1996, Ellis et al.~1996) show that field galaxies have a luminosity
function $n(L)$ far shallower than $n(L) \sim L^{-2}$.  This is normally
reconciled with a steep mass function by arguing that the formation of
stars in low-mass galaxies is inefficient (e.g.~Efstathiou 1992,
White \& Kauffmann 1994, Efstathiou 2000) 
so that low-mass galaxies have far less luminosity
per unit total mass (including dark matter) than high-mass galaxies.  
  
Recent observations have, however, revealed steep luminosity functions
$n(L) \sim L^{-2}$
in both the Virgo (Phillipps et al.~1998) and Fornax (Kambas et al.~2000)
clusters.  Neither of these are particularly rich clusters, like the
Coma cluster, but the important thing is that
they are (unlike Coma)
close enough to us that that their luminosity functions 
can be measured directly down to very faint limits (see Fig.~2 of Trentham
1998).  
On the other hand the luminosity function of galaxies in the Ursa Major
Cluster (Trentham, Tully \& Verheijen 2000), a 
nearby diffuse group of spiral galaxies about
one-twentieth as massive as the Virgo Cluster (Tully et al.~1996), is far
shallower than this. 
The Ursa Major and Virgo/Fornax luminosity functions are
inconsistent at a high level of confidence.  The advent of
mosaic CCDs on large telescopes and the consequent improved depth to
which large areas of the sky can be surveyed to very faint magnitudes
(the Trentham et al.~(2000) Ursa Major survey had a 1$\sigma$ 
surface-brightness threshold
fainter than $\mu_R = 27$ mag arcsec$^{-2}$) means that
the long-standing worry (Disney 1976) of there existing many galaxies
in diffuse environments that are missing from (previously shallow)
surveys because their
surface-brightnesses are too low is far less serious than it used
to be. 

It also appears that the luminosity function of the Local Group
(van den Bergh 2000 and references therein) is not steep between
$M_R = -13$ and $M_R = -11$, which is where the Virgo Cluster
luminosity function is found to be steep; galaxies that have been
discovered recently in the Local Group, like Cetus (Whiting, Hau \&
Irwin 1999) are fainter than this, so that 
van den Bergh's luminosity function is probably complete down to $M_R=-11$. 
The shallow luminosity function
that was seen in Ursa Major may well therefore
be a common feature of diffuse environments. 
This would be in apparent contrast to the upturn in the luminosity
function of dwarf irregular galaxies in the field seen by
Marzke et al.~(1994), but the value of $\alpha \sim -1.8$ quoted
by those authors comes from a Schechter (1976) function fit to the
luminosity function and is only very indirectly related to the faint-end
power-law index of the luminosity function as discussed above (the Schechter 
$\alpha$ is determined primarily by the shape of the bright end of the
luminosity function appropriate the the particular kind of galaxy being
studied;
see Section 2 of Trentham 1997) 
 
Let us assume that the mass function of halos is the same in small galaxy
clusters like Virgo and Fornax as it is in large galaxy groups like Ursa
Major.
This would be likely in any hierarchical clustering model of galaxy
formation in low-density environments.    
It would be less true in very 
high-density environments like the cores of very rich
galaxy clusters like Coma
where dark-matter halos can be tidally disrupted 
(see the high-resolution numerical simulations of Ghigna et al.~1998),
but in this case the variation of the mass function with density is in
the {\bf opposite} sense to what is observed in Virgo, Fornax, and Ursa.
Major -- namely, small galaxies are rarer in the denser environments.   
It follows from this assumption about the mass function that
there must be many galaxies
in the diffuse environments that are completely dark and never formed
stars (see also the discussion in Section 5.2 of Klypin et al~1999).  

There could be many reasons for why this happened, like
gas availability (Virgo and Fornax have substantial X-ray gas halos so
presumably had a reservoir of gas in the past which has since been
gravitationally heated --- Ursa Major does not) and the tidal creation
(Barnes \& Hernquist 1992a) and destruction 
(Putman et al.~1998 and references therein) of dwarfs.
The lack of gas availability for low-mass halos 
in low-density environments may follow from
the heating and ionization of the intergalactic medium 
by OB stars and supernovae over cosmic
time (Klypin et al.~1999; Bullock, Kravstov \& Weinberg  2000). 
Whatever the mechanism, it must 
generate a large effect: there are more than ten
times as many $M_R = -11$ dwarfs in Virgo per $M_R = -20$ giant galaxy
than in Ursa Major (Phillipps et al.~1998, Trentham et al.~2000). 
But the existence of large numbers of dark galaxies in the
diffuse environments does not depend
on the details of this mechanism; it depends only on the assertion that the
mass function does not vary. 
The fact that it 
is also predicted by CDM theory, and the success of that theory at
predicting galaxy properties and large-scale
structure on many different scales (e.g.~Jenkins et al.~1998;
Fontana et al.~1999;
Katz, Hernquist \& Weinberg 1999; Kauffmann et al,~1999)  
simply
gives additional weight to the hypothesis that these dark galaxies exist. 
In CDM theories, these dark halos in low density environments will
tend to have formed during most of the history of the Universe, except
at very early times (then these halos could form stars).
They probably are still forming now.

The dark galaxy hypothesis is one way to solve the problem of
CDM theories overproducing large numbers of small galaxies without having
to abandon the theories and their other successes.  An alternative way to  
perturb the theories so as to avoid this discrepancy is to force  
the cold dark matter to be self-interacting (Spergel \& Steinhardt 2000,
Burkert 2000; however see Miralda-Escud{\'{e}} 2000).     
In this case the mass function of low-mass galaxies in the field
is substantially
shallower than $n(M) \sim M^{-2}$. 

If we accept this possibility of many dark galaxies existing, 
we can then ask the question: 
can we find them, and how?  This is the subject of the
present paper.  The optimum methods of searching for dark galaxies  
depend on the nature of the dark matter, in particular 
whether or not any of the
dark matter is baryonic. 
The results also depend on the total masses of the dark galaxies,
which are unknown.  Given the arguments in the previous paragraphs,
they will be similar to the total masses of the galaxies that
created the upturn in the Virgo luminosity function.  These have
absolute magnitudes $M_R > -13$, and so have stellar masses of $\leq 10^7  
{\rm M}_{\odot}$ (assuming an $R$-band stellar mass-to-light ratio of 1).
The total masses are therefore likely to be $\leq 10^8 {\rm M}_{\odot}$,
given a cosmic ratio of baryonic matter to halo dark matter of 0.1 (see
Section 2 and references therein) and the fact that the fraction of baryonic
matter in galaxies which {\it did} form stars is likely to be close to this
number or lower; the present-day masses of the Virgo dwarfs may be lower
than this due to tidal stripping of their dark matter within the Virgo
cluster but here we are concerned with the initial masses of the halos.  
We therefore restrict the discussion in this paper to
dark galaxies with masses $\leq 10^8 {\rm M}_{\odot}$.  

In Section 2 we review current observations that give
some constraints on the dark matter composition.  
In Section 3 we describe ways that dark galaxies can be  
observed and structure this section according to the possibilities
highlighted in Section 2.  In Section 4 we discuss further where are the
best places in the sky to make the observations described in Section 3.
Throughout this work, we assume the following cosmological parameters: 
$H_0 = 65\,\, {\rm km} {\rm s}^{-1}
{\rm Mpc}^{-1}$,
$\Omega_{\rm matter}=0.3$,
$\Omega_{\Lambda}=0.7$.  
The majority of the observational tests that we suggest
target local dark galaxies and the strategies do not depend significantly
on the cosmology.

\section{Properties of dark galaxies}

\subsection{The cold dark matter component}
It is well established (e.g.~Persic \& Salucci 1988, Pryor \& Kormendy
1990, Mateo et al.~1993) that dwarf galaxies have very large dark
matter fractions.  The smallest galaxies known (Draco and Ursa Minor)
are at least 99 per cent dark matter by mass (Pryor \& Kormendy 1990).  

These very high dark matter fractions probably result from  
baryonic blowout caused by
supernova-driven winds from the few high-mass
stars that formed early in the history of the galaxy
(Dekel \& Silk 1986).  Although the light profiles for very low-luminosity 
galaxies are reasonably well determined, the dark matter profiles are
poorly constrained by observation since the dark matter extends well
beyond the observed stars so that no tracers of the mass are
available (see Pryor \& Kormendy (1990) and Mateo et al.~(1993) for
some models).   
  
So there is very little direct information about the distribution of
dark matter in the smallest galaxies.  Most of the dark matter is
presumably normal cold dark matter, with global ratios of cold to baryonic
matter equal to or less than
the cosmic ratio (about 
one-tenth, from a comparison of
the cosmic total mass density $\Omega \sim 0.3$ e.g.~Perlmutter et al.~1999  
to the baryonic mass density $\Omega_{\rm b} \sim 0.03$
e.g.~Smith et al.~1993). 
The dark galaxies that we study in the current paper are the (more common)
analogues of these low luminosity galaxies, these being the ones that never
made any stars at all.  So the profile of CDM in these completely
dark galaxies is difficult to estimate.  
A reasonable possibility is 
that the CDM 
component will have a density profile
close to a Navarro, Frenk \& White (1996, 1997; hereafter NFW) 
one, assuming that the 
scaling relations derived by
those authors from simulations of much larger bound systems are valid for 
low-mass ones. 
Recall that CDM theory and the observed
absence of low-luminosity galaxies in diffuse environments were the main
motivations for hypothesizing the existence of such galaxies, and these
in conjunction would suggest that a NFW density profile would
be suitable for the cold dark matter, particularly if the baryonic
contribution to the dark matter is small.   

\subsection{The baryonic component}

However, the dark matter
component might not be purely CDM and there might be a 
baryonic component.  If this is true in dark galaxies,
there are important implications for detectability. 

One piece of evidence that baryonic dark matter exists 
is that some 
galaxies have density profiles
inferred from rotation curves that are not consistent with extrapolations of
the NFW profile to lower mass
systems (Moore 1994, Burkert \& Silk 1997, Navarro 
\& Steinmetz 2000). 
The Milky Way, for example, has less mass interior to the solar circle
than predicted from an NFW profile (Navarro \& Steinmetz 2000); 
its density profile is shallower than
$r^{-1}$.
The case of the dwarf spiral DDO 154 is studied in particular detail by
Burkert \& Silk (1997). 
As these authors point out, a high baryonic dark matter fraction 
(50 per cent, far in excess of the global value of 6.5 per cent derived
by Hernandez \& Gilmore 1998)
can alleviate this problem.  Another way to solve the problem is to
force the CDM to be self-interacting, as outlined in the previous
section (Spergel \& Steinhardt 2000).
However, systematic observational effects (e.g.~beam smearing) 
may themselves be the source the problem.  Indeed, van den Bosch \&
Swaters (2000) show that when the effects of beam smearing and the physics
of adiabatic contraction are taken into account, the mass models
permitted by the rotation curves
of dwarf spiral galaxies are consistent with CDM theory (see also the
discussion in Flores \& Primack 1994). 

Another piece of evidence that baryonic dark matter exists
is the cosmic baryonic budget 
of Fukugita, Hogan \& Peebles (1998).  These authors cannot account for
more than about 20 per cent of baryons in the Universe (as inferred from
Big Bang nucleosynthesis) in stars and gas.  Some of the 
missing baryons may be in
a form that is dark, and 
could account for some fraction of
the dark matter in the kind of halos we are looking for.
If this is the case, however, dark matter that resides in small galaxies
probably cannot form a major part of the missing baryons, since the 
cosmological density residing in baryons in the high-redshift
intergalactic medium is close to the nucleosynthesis value (Weinberg
et al.~1997), and it is difficult to see how hot gas like this can
condense into the small halos, particularly in light of the rationale 
in Section 1 as to why the galaxies are dark.  However, some small
fraction of the missing baryons could certainly be associated with
small CDM halos. 

In the rest of this section we consider some possible forms which this
baryonic dark matter 
might take.  

\subsubsection{Brown dwarfs}

Brown dwarfs are stars with masses lower than the hydrogen-burning limit of
0.08 M$_{\odot}$.
They are not a significant contributor to the mass in young star-forming
regions like the Pleiades (Hodgkin \& Jameson 2000).
In addition, they are probably not a major component of the
dark mass in the outer parts of the 
Milky Way, as inferred from the MACHO (Alcock et al.~2000) 
and EROS (Afonso et al.~1998) 
microlensing experiments.  In addition to generating microlensing effects,
brown dwarfs emit thermal radiation at
mid-infrared wavelengths and can be detected this way.  
In the external spiral galaxy
NGC 4565 (Beichman et al.~1999), there are constraints
from {\it ISO}, but they do not rule out the possibility of a significant
fraction of dark matter 
being brown dwarfs.  
In dwarf spiral 
galaxies the {\it ISO} measurements (Gilmore \& Unavane
1998) are also unable to rule out
a significant
fraction of dark matter (inferred to exist from rotation curves) 
from being in the form of
brown dwarfs.  For a review of the plausibility of
brown dwarfs as dark matter, see Gilmore (1999). 

In the lowest mass galaxies, many of which may be completely dark,
let us hypothesize that some of the dark matter
may be brown dwarfs. 
Even if the fraction of the dark matter in this form only is a few per cent,
this hypothesis
would require that they formed with a very different stellar mass
function (IMF) than that appropriate for the solar neighbourhood (e.g.~Kroupa, 
Tout \& Gilmore (1993).
This discrepancy follows from the
paucity of normal stars in dark galaxies.
This assertion is somewhat unattractive because the
solar-neighbourhood IMF seems be universal
wherever direct observations may be made, in both Population I and Population
II environments 
(Gilmore \& Howell 1998).  
On the other hand, the brown dwarfs that we consider here may be
Population III objects, and
a different IMF for Population III stars could result from a variety
of physical processes e.g.~the molecular hydrogen out of which 
the stars formed
was generated in a different way, probably by triple-H collisions
(Palla, Salpeter \& Stahler 1983), since
no dust grains existed (adsoprtion of H atoms onto dust is the most
efficient way of making molecular hydrogen in Population I and probably
Population II environments).

\subsubsection{Molecular hydrogen}
Molecular hydrogen (H$_2$), if pristine and not contaminated with dust (which
can be seen with submillimetre bolometers and CO line receivers) is 
notoriously difficult to detect (Combes \& Pfenniger 1997).  
Pfenniger, Combes \& Martinet (1994) argue that
unenriched molecular gas could be a common form of halo dark matter in
giant galaxies.  
The main arguments against these clouds being the predominant form
of halo dark matter are mainly cosmological: (i) the cosmological
density of dark matter is
a factor of several higher than the cosmological
density of baryons, as outlined in Section
2.1, and (ii) these clouds
cannot survive at high redshift where the temperature of the microwave
background is higher (Wardle \& Walker 1999), so their formation has to
be finely tuned to happen at low redshift, well after the bulk of star
formation in the Universe (Madau et al.~1996). 
But there is nothing to suggest that small clumps of molecular hydrogen
cannot exist as trace amounts of dark matter in late-forming galaxies. 
Such clouds are thermally stable (Wardle \& Walker 1999).
If such clouds exist in
dark galaxies, they might help the galaxies be detected, for example
by increasing
the angular extent of the galaxy that has a surface density higher than
the critical density for strong lensing (see Section 3.1). 

\subsubsection{Stellar remnants}
Two kinds of stellar remnants could contribute at some level to the
dark matter: compact objects like neutron stars and black holes, and
cold white dwarfs. 

Early supernovae will leave remnants with kick velocities in excess of
the escape velocities of the galaxies.
Pulsar birth speeds in the Galaxy are typically in excess of
200 km s$^{-1}$ (Lyne \& Lorimer 1994, Hansen \& Phinney 1997).  
The escape velocities of the small galaxies which are hypothesized to
form stars in the very early Universe are small ($< 200$ km s$^{-1}$)
since big galaxies with deep gravitational potential wells
have not yet had time to form (see Tegmark et al.~1997
for a description of the formation of the very first galaxies
and the relevant mass scales, and
Gnedin \& Ostriker 1997 for a cosmological simulation of the early formation
of galaxies; note that the escape velocity of a virialized galaxy is
approximately twice the velocity dispersion -- e.g.~p.~119 of Saslaw
2000). 
If the physics of core-collapse supernovae is invariant with cosmic time,
early-forming
remnants will therefore leave the parent galaxies and intergalactic space
may contain many such objects.  It is therefore possible that later-forming
dark matter halos of the type described in Section 1 may accrete a handful
of these remnants.  They could therefore be present in trace amounts in
dark galaxies.  

One interesting recent result is the detection of three cold
($< 4000$ K), and therefore
extremely old, white dwarfs (CWDs) in the Milky Way halo (Hambly, Smartt \&
Hodgkin 1997, Harris et al.~1999, Ibata et al.~2000).
These have spectral energy distributions that peak at  
about 1 $\mu$m (Hodgkin et al.~2000) due to 
molecular collision-induced absorption at longer wavelengths (Hansen 1999). 
They are, however, very faint ($M_V \sim 17$, $V$-band mass-to-light
ration = $5 \times 10^{4}$ in solar units) and so are almost
dark.  The presence of even 
these three objects given the selection effects appropriate to the
above surveys implies the existence of
massive progenitors that are sufficiently numerous that they are not
consistent with an extrapolation of the normal halo population to higher
masses, given a normal stellar IMF like that of Kroupa et al.~(1993). 
These progenitors could be Population III objects.  
No abundances are known for the CWDs, since no metals are
present in their atmospheres; the metallicites of their
progenitors are therefore
unconstrained. 
The CWDs may be associated with 
very early formation of the Milky Way
(i.e.~at times before the normal halo population and globular
clusters formed).
Alternatively they could have been accreted by the Milky Way halo from
intergalactic space if they formed in fragile small galaxies in the early
Universe which had been tidally disrupted.  The latter is suggested
by the fact that a substantial population of old white dwarfs forming
{\it within} the Milky Way halo would have resulted in too much 
heavy element production  (see Gilmore \& Unavane 1998 and
references therein; also Gilmore
1999).  The comments made in Section 2.2.1 about the plausibility of
Population III having an IMF that is different from Populations I and II
is valid here too, although it would be in the opposite sense to the
difference hypothesized in that section i.e.~skewed towards high, not low,
masses.  

\section{Techniques for finding dark galaxies}

\subsection{Gravitational lensing}

In this section we assess under what conditions a dark galaxy could be 
detected through gravitational lensing of background sources.

\subsubsection{Strong lensing by dark halos}

\begin{figure*}
\begin{minipage}{170mm}
\begin{center}
{\vskip-3.5cm}
\psfig{file=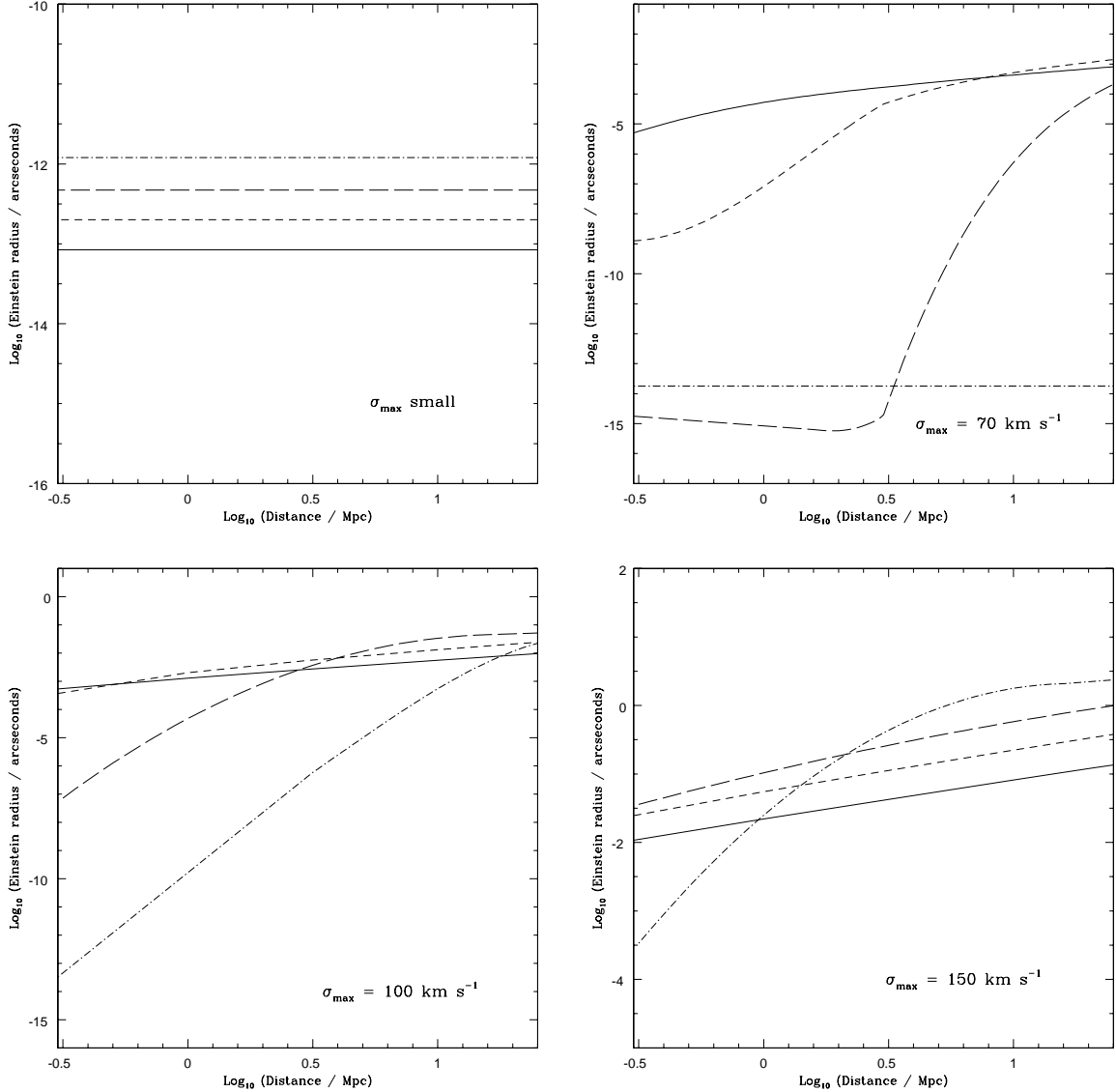, width=18.65cm}
\end{center}
{\vskip-5.7cm}
\caption{Einstein radii for NFW lenses at various distances.  The NFW profiles
are chosen according to their virial masses $M_V$ and maximum velocity
dispersion $\sigma_{\rm max}$.  In all panels the different lines represent
lenses of different mass: $M_V = 10^7 {\rm M}_{\odot}$ -- solid line;
$M_V = 10^8 {\rm M}_{\odot}$ -- short dashed line;
$M_V = 10^9 {\rm M}_{\odot}$ -- long dashed line;
$M_V = 10^{10} {\rm M}_{\odot}$ -- dotted-dashed line.
The different panels are for halos of different $\sigma_{\rm max}$, as
shown.  By ``small'' $\sigma_{\rm max}$ we mean 
1.9 km s$^{-1}$ ($M = 10^7 {\rm M}_{\odot}$), 
4.1 km s$^{-1}$ ($M = 10^8 {\rm M}_{\odot}$),
8.6 km s$^{-1}$ ($M = 10^9 {\rm M}_{\odot}$),
or 18.1 km s$^{-1}$ ($M = 10^{10} {\rm M}_{\odot}$).
}
\end{minipage}
\end{figure*}

Following Section 2.1, let us assume that the dark matter density has a NFW
(Navarro, Frenk \& White 1996, 1997) profile:
\begin{equation}
\rho(r)=\frac{\rho_0}{(r/r_{\mathrm{s}})(1+r/r_{\mathrm{s}})^2}
\end{equation}
where the scale density $\rho_0$ and radius $r_s$ are determined
uniquely by two free parameters of the halo, in the current work
chosen to be the virial mass $M_V$ and
maximum velocity dispersion $\sigma_{\rm max}$.  
If we fix $M_V$ and $\sigma_{\rm max}$, then $\rho_0$,  $r_s$   
and a concentration index $C$ are uniquely defined by the equations
\begin{eqnarray}
\lefteqn{\nonumber
\rho_0=3.6\times10^8 \, \left(\frac{\sigma_
{\mathrm{max}}}{100\,
\mathrm{km\,s^{-1}}}\right)^6\left(\frac{M_{\mathrm{v}}}{10^{11
}\,M_{\odot}}\right)^{-2} \times } \\
& \>\>\>\>\>\>\>\>\>\>\>\>\>\>\>\>\>\>\>\>\>\>\>\>\>\>\>\>\>\>\>\>
{(\ln(1+C)-C/(1+C))^2 \,\, 
{{\rm M}_{\odot}\,\mathrm{kpc^{-3}}; 
}}
\end{eqnarray}
\begin{eqnarray}
\lefteqn{\nonumber
r_{\mathrm{s}}=2.8 \,\left(\frac{M_{\mathrm{v}}}{10^{11}\,M_
{\odot}}\right)\left(\frac{\sigma_{\mathrm{max}}}{100\,\mathrm{km\,s^{-1}}}
\right)^2 \times} \\ 
&  \>\>\>\>\>\>\>\>\>\>\>\>\>\>
     \>\>\>\>\>\>\>\>\>\>\>\>\>\>\>\>\>\>\>\>\>\>\>
(\ln(1+C)-C/(1+C))^{-1} \,\, {\rm kpc};
\end{eqnarray}
\begin{equation}
r_{\mathrm{s}}=\frac{1}{C}\left(\frac{3M_{\mathrm{v}}}{800\rho_{\mathrm{c}}\pi
}\right)^{1/3}.
\end{equation}
Here $\rho_{\rm c}$ is the critical
density of the Universe (throughout this section we assume the redshift $z=0$). 

The radius of the tangential critical curve of an axially symmetric
lens 
(the Einstein radius) 
is equal to the radial distance at which the mean surface density
within that radius equals the critical surface density for strong
lensing 
(Schneider, Ehlers \& Falco 1992, p.~233). 
The mean surface mass density $\bar{\Sigma}$ at 
distances $r=r_{\mathrm{s}}x$ is given by 
\begin{equation}
\bar{\Sigma}=4r_{\mathrm{s}}\rho_0\times f(x),
\end{equation}
where the function
\[
f(x)=\left\{ \begin{array}{l@{\quad : \quad}l} 
\frac{1}{x^2}\left(\frac{2}{\sqrt{1-x^2}}
\,{\rm arctanh}{\sqrt{\frac{1-x}{1+x}}}+\ln(x/
2)\right)
& x<1 \\
1+\ln(1/2) & x=1 \\
\frac{1}{x^2}\left(\frac{2}{\sqrt{x^2-1}}\,{\rm 
arctan}{\sqrt{\frac{x-1}{1+x}}}+\ln(x/2
)\right)
& x>1 
\end{array}\right.
\]
The Einstein radius is then the radius of that
circular region in which $\bar{\Sigma}=
\Sigma_{\mathrm{c}}(z_{\mathrm{l}},z_{\mathrm{s}})$ where
$\Sigma_{\mathrm{c}}$ is the critical density for lens redshift
$z_{\mathrm{l}}$ and source redshift $z_{\mathrm{s}}$.
For nearby lenses at distance $D$
such that $z \ll 1$, and sources at cosmological distances,
the critical density is
\begin{equation}
\Sigma_c = { {c^2}\over {4 \pi G D}}
\end{equation}
where $c$ is the speed of light and $G$ is the gravitational
constant.  If we now set $M_V$ and $\sigma_{\rm max}$ for NFW lenses
and place them at distance $D$, we can work out the Einstein radii for
these lenses.  This is done in Figure 1. 
 
Radio VLBI interferometry can achieve an angular resolution of about
$10^{-4}$ arcsec.  This is the best resolution available for
radioastronomy until space interferometers are deployed.
From Fig.~1, dark galaxies with $\sigma_{\rm max} > 70 \, {\rm km} \, 
{\rm s}^{-1}$ at distances greater than 10 Mpc
could strongly lense background quasars and generate
multiple images with separations that could be detected by VLBI
interferometry.  The number of such objects that we might expect to find
depends on the fraction of low-mass dark galaxies with
$\sigma_{\rm max} > 70 \, {\rm km} \,
{\rm s}^{-1}$.  Unfortunately this velocity is high, larger than the
maximum velocity dispersion of dwarf spiral galaxies like DDO 154
(Burkert \& Silk 1997) and much larger than the velocity dispersions of
the dwarf elliptical galaxies studied by Kormendy (1987), thought to
be the counterparts of the dark galaxies which {\it did} make stars.
It is therefore not surprising that to date
no lens system where the lens is dark has been detected with VLBI. 

The detectability of dark galaxies through 
strong gravitational lensing becomes
much higher if some fraction of the dark matter is very compact and
does not follow a NFW profile, say in
a supermassive black hole at the galaxy center.  
For a point mass $M$ at distance $D$, the Einstein radius for sources
at cosmological distances is
\begin{equation}
\theta_{\rm E} = 0.029 \, \left( { {M}\over{10^6 {\rm M}_{\odot}}}\right)^{0.5}
\,
\left( { {D}\over{10 \,{\rm Mpc}}}\right)^{-0.5} \,\,{\rm arcsec}, 
\end{equation}
which is 0.029 arcseconds 
for masses of 10$^6$ M$_{\odot}$, easily detectable with 
current radio telescopes.  There is, however, no evidence that low-luminosity
late-type galaxies (the visible counterparts of the dark galaxies) have
central mass concentrations this large (Kormendy \& Richstone 1995). 
We therefore do not expect
the kind of dark galaxies that we hypothesize here to have them either. 
If the NFW density profile can indeed by extrapolated to low masses for
galaxies composed almost entirely of CDM,
the only circumstances where this might be a realistic possibility is
if dark galaxies have dense baryonic cores (for example, these could be
made of molecular gas that somehow
never made stars, as argued in Section 2.2.2).

\subsubsection{Weak lensing by dark halos}

The previous analysis has concentrated on strong lensing.  But if dark
halos are present in enough number, their cumulative weak lensing effects
might be measurable.

In a recent paper,
Natarajan \& Kneib (1997) showed how the granularity in the shear map of
a galaxy cluster could be used to infer the presence of dark halos around
individual galaxies.  With current data, this method can only be used to look
for massive ($> 10^{10}\,{\rm M_{\odot}}$) halos in distant, rich clusters
with small angular extent on the sky, due to shot
noise limitations.  However, if it is possible at some
time in the future to image very deeply over large angular areas of
the sky at high resolution (say, with an array of telescopes using
orthogonal transfer CCDs; Kaiser, Tonry \& Luppino 2000), then   
it should be possible to apply the analysis of Natarajan \& Kneib to
more diffuse, spiral-rich groups of galaxies to look for massive dark
halos.  These environments would be particularly suitable for such an
analysis since (i) they are dynamically unevolved, and so likely to have
many dark halos based on the arguments in Section 1, and (ii) they do
not have substantial dense dark matter concentrations associated with the
overall structure as opposed to individual galaxies (as do rich clusters),
small pieces of which could be tidally torn off and mimic dark galaxies. 

\subsubsection{Strong lensing by individual brown dwarfs or stellar remnants}

We have argued in the previous section
that it is possible that dark galaxies may have some
brown dwarfs or stellar remnants associated with them.  These 
could be responsible for microlensing events, like those of Magellanic
Cloud stars observed by the MACHO and EROS collaborations.
A large number of such events from a small angular region of the sky
may indicate the presence of an anomalously large number of
brown dwarfs 
or stellar remnants that are very close in physical space.  If these
are not so close as to be self-gravitating, they could signify the
existence of a dark satellite somewhere between us an the Magellanic Cloud.  
See Widrow \& Dubinski (1999) for a detailed simulation-based treatment
of this phenomenon. 

The non-detection of such a phenomenon so far places a limit on the joint
density of dark satellites around the Milky Way and the contribution of
brown dwarfs and stellar remnants to their total masses.  Unfortunately, the
angular area of the sky probed by the above experiments 
combined with the predicted
number density of dark satellites (about 10$^3$ in the
entire Local Group; see Klypin et al.~1999 -- the volume of the cone
subtended by the Magellanic Cloud to an observer on Earth is only
about 10$^{-7}$ of the volume of the Local Group) 
means that this constraint is not strong.

\subsection{Infrared radiation}

\subsubsection{Mid-infrared Thermal Radiation from Brown Dwarfs}

If a significant fraction of the dark matter in dark galaxies is in the
form of brown dwarfs, as highlighted in Section 2.2.1, these
may be detectable at mid-infrared (5  $\mu$m $-$ 30  $\mu$m)
wavelengths.  They therefore may offer one potential way to directly see
an otherwise dark galaxy.
Gilmore \& Unavane (1998) placed upper limits on the
number of brown dwarfs in a sample of dwarf spiral galaxies using 
$ISO$ observations at 7 $\mu$m and 15 $\mu$m.  With the next
generation of mid-infrared space observatories and infrared
cameras, like $SIRTF$-MIPS, 
far more sensitive observations will be possible.  The following
quantitative analysis is based on the strategy adopted by
Gilmore \& Unavane for their $ISO$ observations. 

\begin{figure}
\begin{center}
\vskip-4mm
\psfig{file=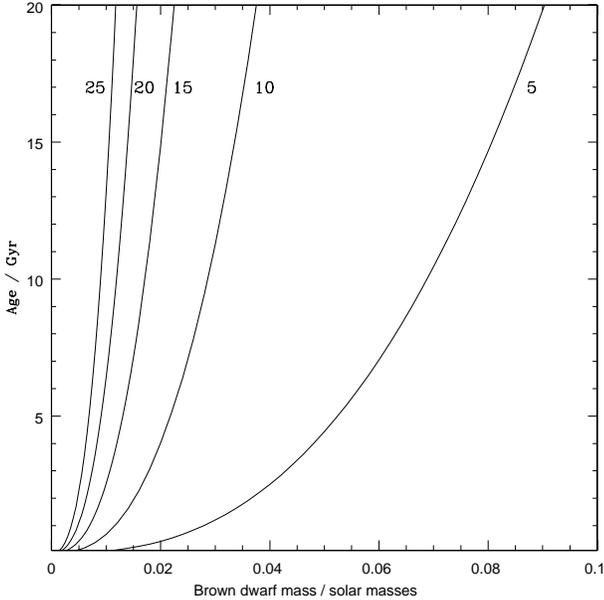, width=8.65cm}
\end{center}
\vskip-3mm
\caption{
Contours of wavelength in $\mu$m at which the maximum
flux density is emitted for brown dwarfs of various ages and
temperatures.  The scaling relations of Stevenson (1986) are
used and the brown dwarfs are assumed to radiate as blackbodies,
as is appropriate for low
metallicities given the more detailed analyses of
Burrows et al.~(1997) and Baraffe \& Allard (1997; see the
discussion in Section 6.2.5 of Gilmore \& Unavane 1998).
}
\end{figure}

Brown dwarfs have spectral energy distributions that peak at mid-infrared
wavelengths, since (assuming they have low metallicities as would be 
appropriate for early-forming, or Population III objects) they radiate
as cool blackbodies (see Figure 2). 
A dark galaxy may consist of a population of such objects (in the analysis
below we assume that they all the same mass and age).
For a spherical brown dwarf population with
projected density profile $\Sigma (\theta)$, 
the flux density
emitted per unit angular area  
from an element at angular distance $\theta$ from the center is
\begin{eqnarray}
\lefteqn{\nonumber
I_{\nu} = 
0.0145 \,
M_8 \, \nu_{14}^3 \,
t^{-0.01} \,
m^{-2.66} \,
D_{1}^{-2} \,
\times} \\ 
& \>\>
(\exp{\left( 0.278 \, \nu_{14} \,
t^{0.31} \, m^{-0.79} \right)} - 1)^{-1} 
\, I(\theta)
\, \, {\rm mJy} \, {\rm rad}^{-2},
\end{eqnarray}
where
\begin{equation}
I({\theta}) = { {\Sigma(\theta)}\over{\int_{0}^{\infty}
     \Sigma(\theta) {\rm d}^{2}{\theta}} }.
\end{equation} 
Here 
$m$ is the mass of the brown dwarfs in ${\rm M}_{\odot}$, 
$t$ is the age of the population in units of Gyr,
$M_{8}$ is the total population mass in units of
  $10^8 \, {\rm M}_{\odot}$, $D_1$ is the distance to the galaxy in
units of 1 Mpc and 
$\nu_{14}$ is the frequency in units of $10^{14}$ Hz. 
The scaling relations of Stevenson (1986) are assumed. 

Whether or not such a population of brown dwarfs
is detectable with an instrument like $SIRTF$-MIPS
(http://ipac.caltedu.edu/sirtf/)
depends on two issues: (i) the instrument sensitivity, and
(ii) the flat-fielding capability.   
Suppose we want to look for a dark galaxy 
at a distance of 1 Mpc 
consisting of $10^8 {\rm M}_{\odot}$
(i.e.~10 per cent of a galaxy with total mass $10^9 {\rm M}_{\odot}$) of
low-metallicity brown dwarfs,
each having mass $0.01 {\rm M}_{\odot}$ and age 15 Gyr.   
From Fig.~2, the optimal wavelength to observe is about 25 $\mu$m, so a
suitable instrument to use is $SIRTF$-MIPS with a 24-$\mu$m
filter.  For this instrumental configuration, the 
beam area is 0.01 arcmin$^2$ (Blain 1999).   
From Equation XXX, the total flux in the galaxy seen through this
filter is 0.23 mJy.
If we assume a projected density
profile for the brown dwarfs that is exponential with scale-length 0.1 kpc
(this is the scale-length of visibile matter in the smallest dwarf
elliptical studied by Kormendy 1987), the total flux in a beam centered
on the galaxy center is 0.013 mJy.    
The noise equivalent flux density (NEFD) for this instrumental configuration is
1.8 mJy$\,$Hz$^{-1/2}$ (Blain 1999), so a 3$\sigma$ detection could be
obtained with an integration time of 200 ks, or about 2.3 days.
Probably a somewhat lower integration time can be used if we sum the
contributions from several beams centered on the galaxy center; the optimal
strategy will depend on the detailed density profile.
The second issue we need to consider is the flat-fielding capability of
the instrument.  For $SIRTF$-MIPS operating at 24 $\mu$m,
the field of view is about 5 arcminutes $\times$ 5 arcminutes (Blain 1999),
about 14.5 scale-lengths for the object above.  
The numbers above assume perfect flatfielding, and systematic variations
in the sensitivity of the detector would need to be accurately corrected
for if such a measurement is to be feasible.
This would be even more important if we tried to
identify dwarfs with $D_1 << 1$ this way.

The integration time required to achieve a given signal-to-noise
scales as $I_\nu^{-2} \sim M_8^{-2}$, so less 
substantial populations of these brown dwarfs could could be only detected 
with $SIRTF$-MIPS if very long integrations are used.
Future generations of observatories will, however,  
have lower NEFDs, as instrument
sensitivities improve, and this kind of project becomes feasible even
for very trace populations.
Dark galaxies would be particularly straightforward objects to observe
this way since there is no contamination from long-wavelength emission from
normal stars and their debris. 

\subsubsection{Near-Infrared Radiation from Cold White Dwarfs}

\begin{figure}
\begin{center}
\vskip-4mm
\psfig{file=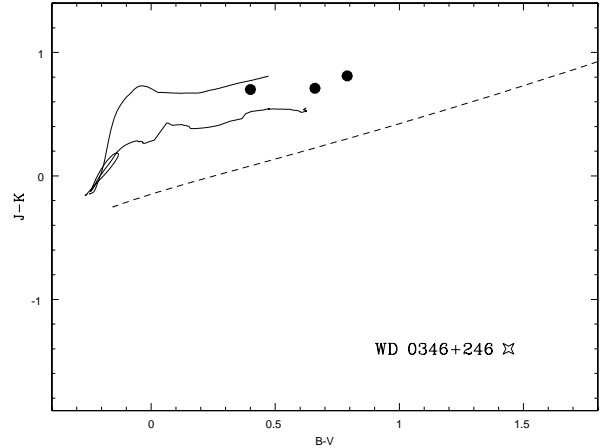, angle=-90, width=8.65cm}
\end{center}
\vskip-3mm
\caption{
A $J-K$ vs.~$B-V$ colour-colour diagram 
showing the location of WD 0346+246 (the star)
relative to various low
surface-brightness stellar systems (the filled circles) 
that might mimic a diffuse population of cold white
dwarfs in an otherwise dark galaxy.  These are (in order of
increasing $B-V$): ESO 462$-$36 (a blue low surface-brightness
star-forming galaxy), NGC 4472 dw10
(a blue Virgo dwarf elliptical galaxy) and NGC 4472 dw8
(a red Virgo dwarf elliptical galaxy).
The photometry is from Hodgkin et al.~(2000) for
WD 0346+246, from Bergvall et al.~(1999) for
ESO 462$-$36 and from Caldwell (1983)
and Bothun \& Caldwell (1984) for the Virgo dwarfs.
The thin solid line represents the colours 
for a stellar population generated in an instantaneous
starburst seen at times 10$^5$ to $2 \times 10^9$ years
after the burst, assuming a Miller-Scalo stellar IMF
and a metallicity of 1/50
solar, from the models of Bruzual \& Charlot (1993). 
This would be appropriate for a pristine star-forming
HII galaxy like I Zw 18 (e.g.~Hunter \& Thronson 1995).
The thick solid line represents the colours 
for a continuously-forming stellar population of solar
metallicity, from the same models. 
The dashed line represents a blackbody.
}
\end{figure}

Cold white dwarfs have spectral energy distributions that peak at about
1 $\mu$m (Hodgkin et al.~2000) so that a population of such
objects  could be detected as a faint excess
above the sky in $J$-band (1.25 $\mu$m).  Such a population
could readily be distinguished
from other low surface-brightness stellar systems on the basis of their
optical+near-infrared colors (see Figure 3). 
Imagine now an object containing 10$^{8}$ M${\odot}$ of CWDs with an
exponential mass profile with scale-length $h$. 
If placed at a distance of 1 Mpc,
such an object will have a $J$-band surface-brightness profile 
\begin{eqnarray}
\lefteqn{\nonumber
\mu_J (\theta) = 33.47 + 5 \,\,{\rm log}_{10} 
\left( {{h}\over{1 \, {\rm kpc}}}\right)
}\\
& \>\>\>\>\>\>\>\>\>\>\>\>\>\>\>
+ 0.00525 \left( {\theta}\over{1 \,{\rm arcsec}}\right)
\left( {{h}\over{1 \, {\rm kpc}}}\right)^{-1}
\,{\rm mag} \,{\rm arcsec}^{-2},  
\end{eqnarray}
which is everywhere 
much less than the sky surface-brightness in $J$-band (about
16 $J$ mag arcsec$^{-2}$; Tokunaga 2000). 
This calculation is based on the spectral energy distribution of WD 0346+0246,
which has an absolute $J$-band magnitude of $M_J = 17.60 - 5 \log_{10} 
(28/10) = 15.36$ and a mass of 0.65 M${\odot}$ (Hodgkin et al.~2000).  

Such an object would require very long integration times to detect 
with current telescopes and instruments. 
For example, in order to get a 3$\sigma$ detection using the INGRID
imaging camera on the 4.2 m William Herschel Telescope 
(http://www.ast.cam.ac.uk/ING/Astronomy/instruments/ ingrid/index.html), 
an integration
of 24 ks (about 0.3 days) is required if $h = 0.01$ kpc.  
Additionally, since we are looking for such faint, extended objects, it is
important that the images be flatfielded to very high accuracy.
Integration times scale
as $h^4$, so more diffuse populations will be much more difficult to
detect.  For objects much fainter than the sky, integration times scale
as telescope diameter $D^{-2}$, so going to larger telescopes helps a little,
but very long integrations indeed are still required if  
$h$ is large.  Since part of the motivation for suggesting that some
of the dark matter is baryonic in the first place 
was that luminous dwarf galaxies 
have constant density cores (i.e.~large $h$), this would seem to suggest
that looking for populations of CWDs like this will not be productive.
It is only if they have very centrally peaked mass profiles that we will
see them, but if such an object is detected we have the advantage
of unambiguously knowing what it is given the very substantial difference
in colors between WD0346+246 and other stellar systems, as shown on Figure 2. 

\subsection{Gamma-ray burst afterglows and host galaxies} 
One of the favored theories for the formation of gamma ray bursts (GRBs)
is the merger of compact stellar remnants like neutron stars or black
holes (Narayan, Paczy\'nski \& Piran 1992).  Therefore if two of the
remnants (accreted by a single dark galaxy as described in Section 2.2.3) 
happen to merge,
then a GRB may result.

The interaction between the burst and the surrounding material results
in an afterglow at lower energies (M\'esz\'aros and Rees 1998, 
B\"ottcher et al. 1999).  The properties of this afterglow depend on the
density of the ambient medium, which would be far lower in a dark galaxy
than in a normal star-forming galaxy, by several orders of magnitude.  Thus 
in principle a GRB occurring in a dark galaxy could be identified as
follows.
The beginning of the afterglow phase (Vietri 2000) occurs a time $t_d$ 
after the end of the $\gamma$-ray emission, where  
\begin{equation}
\label{td}
t_d \sim 15\; 
\left(\frac{E}{10^{53}\,{\rm erg}}\right)^{1/3}
\left(\frac{n}{1\, {\rm cm}^{-3}}\right)^{-1/3}
\left(\frac{\Gamma}{300}\right)^{-8/3}\;\,{\rm s}.
\end{equation}
Here $n$ is the density of surrounding matter, $E$ is the total
explosion energy of the burst and $\Gamma$ the initial bulk
Lorentz factor.
In dark galaxies $n$ is unlikely to be much more than the intergalactic
medium value, but in normal galaxies it will probably be close to the
the value appropriate to star-forming regions, particularly if most GRBs
arise from collapsars (Woosley 1993, Paczy\'nski 1998).  Hence GRBs with very
large values of $t_d$ might suggest that the host galaxy is dark.  
The dependence of $t_d$ on $n \Gamma^8$ suggests that any  
difference in $t_d$ due to $n$ can be 
mimicked by a small change in $\Gamma$.  
But the plausible range of $\Gamma$ ($\Gamma > 30$ -- M\'esz\'aros, Laguna
\& Rees 1993 -- and $\Gamma < 1000$ -- Ramirez-Ruiz \& Fenimore 2000) is
so small compared with the plausible range of $n$ ($n \leq 10^{-7} 
\,{\rm cm}^{-3}$ in the IGM, assuming a maximal value from Big Bang
Nucleosynthesis -- Smith et al.~(1993) --- and $n > 300 \,{\rm cm}^{-3}$ in
star-forming molecular clouds -- Sanders, Scoville \& Solomon
1985) that the two
parameters may be individually determined by considering the 
peak frequency (Wijers et al.~1999)   
\begin{equation}
\label{num}
\nu_m\sim3\times10^{15}\,\left(\frac{1+z}{2}\right)^{1/2} 
({E \over 10^{53}\,{\rm erg}}) \,\,t_{\rm day}^{-3/2}
\left(\frac{15\;s}{t_d}\right)^{3/2}\,{\rm Hz}\;
\end{equation}
and intensity at $\nu_m$ (Waxman 1997a,b, Vietri 2000)  
\begin{equation}
\label{fnum}
F(\nu_m) \sim 10
\, C(z) 
\left( {n \over {1 \,{\rm cm}^{-3}}}\right)^{1/2}  
\left( {E \over 10^{53}\, {\rm erg}} 
\right)^{1/2}\,{\rm mJy}
\end{equation} 
of the afterglow, where 
$t_{\rm day}$ 
is the time of observations in units of days since the onset of  
the afterglow and $C(z)$ is given by
\begin{equation}
C(z) =  \frac{2}{1+z}
\left(\frac{1-1/\sqrt{2}}{1-1/\sqrt{1+z}}
\right)^2. 
\end{equation}
Here $z$ is the redshift of the burst.
A large number of GRBs whose afterglows suggest values of $n$ appropriate
to the intergalactic medium value might suggest a substantial population of
dark galaxies.  No host galaxy should be detectable from any of these
events, even with the next generation of extremely large telescopes or the
Next Generation Space Telescope (NGST; http://www.ngst.stsci.edu/).  

\subsection{Atomic hydrogen searches}

Dark halos may over cosmic time have collected small amounts of gas which
is cold and in atomic (HI) form today.   If this HI gas can be detected,
and sufficient numbers of detections are made,
it might signify the existence of dark galaxies.  

One possibility is that these kinds of objects are the low-stellar-mass
counterparts of Local Group galaxies, like LGS3 (Hodge 1994 and
references therein).  That LGS3 was
detected follows from the fact that it was able to turn a small amount of its
gas into stars.  Other systems which did not do this (for whatever reason,
perhaps because they have very low gas masses indeed) would be optically
dark.

Another possibility which has attracted much recent attention (see Section 5.1
of Klypin et al.~1999) is that the high velocity clouds
(HVCs) in the Local Group described by
Wakker \& von Woerden 1997 and Blitz et al.~1999 might be the observational 
counterparts of the dark galaxies predicted by CDM theory. 
The dark halos that happen to lie
in those environments where there is plenty of gas
available could accrete substantial amounts of this gas.  This notion is  
consistent with the observed phenomena that
(i) HVCs tend to be found concentrated
around the Magellanic Stream where there certainly {\it is} a substantial
gas reservoir, and (ii) the gas in HVCs is reasonably enriched (0.1 solar;
Wakker et al.~1999),
and must have experienced chemical evolution somewhere at some
time in the past. 
While intriguing, the interpretation of the HVCs as the observational
counterparts of the dark galaxies is still premature.  As Klypin et al.~(1999)
point out in the last paragraph of their Section 5.1, this assertion relies on
many observational parameters yet to be determined, 
but whose measurement is feasible
with current technology (e.g.~Gibson \& Wakker 1999).  
Two issues that seem particularly important
to establish are 
\vskip 1pt \noindent
(1) that many HVCs can exist at large distance from the Milky Way, so that
they are not just a local phenomenon, and
\vskip 1pt \noindent
(2) the radial metallicity distribution.  If HVCs far away
from the Milky Way are found to have 
unenriched gas that is likely pristine (like I Zw 18, or maybe even more
unenriched), this would
also support the picture that there exist a large number of
primordial HVCs unassociated with larger galaxies.

More generally, it seems a good idea to search for small HI systems.
Unfortunately, if the column densities of such systems are very low,
they cannot be seen in emission without very long integrations due to
the noise properties inherent in HI measurements (Disney \& Banks 1997;
note that the column density threshold of a measurement
$N({\rm HI}) \sim T_s \sqrt{{\Delta V}/t}$, where $T_s$ is the system
temperature, $\Delta V$ is the HI linewidth and $t$ is the integration time,
is independent of the size of the telescope used). 
Very low column density systems
may, however, be detectable as absorption-line systems in  
the ultraviolet
spectra of low-redshift quasars or even stars.  For
the lowest column density systems, extremely high signal-to-noise 
spectra are required; this could be an important application of NGST.  

\subsection{Anomalous stellar associations}

It is also possible that late-forming dark halos may accrete normal stars
from intergalactic space that could have been released during tidal
encounters (Barnes \& Hernquist 1992b) between galaxies.  

If a small number of stars can be identified that are all at the same
distance, are very close in physical space, have a high velocity dispersion, 
and have very different inferred
ages and metallicities, this might suggest that we are seeing a
group of stars held together by a dark halo that has collected these
stars over cosmic time.   It would
be  difficult to explain these properties in
concordance in any other way.  Certainly if such objects are ever found in
appreciable number, this would be evidence for a dark galaxy population.

Observationally, the most difficult aspect of 
this problem will be establishing
stellar distances.  Let us assume that the faintest stars in an
anomalous stellar association are main-sequence stars with
ages that are a reasonable fraction of
a Hubble time.
Most such stars have absolute magnitudes close to that 
of the Sun i.e.~$M_B = 5.5$.
Currently the most precisely-determined distances come
from parallax measurements with the Hipparcos satellite
(http://astro.estec.esa.nl/Hipparcos/), which can determine parallaxes for
stars with $B \leq 9$ to an accuracy of about 0.002 arcseconds.      
This satellite could therefore only be used to find such associations 
within 50 pc or so of us i.e.~very nearby. 
Associations more distant than this will have stars too faint to allow
accurate parallax measurements, so their distances cannot be determined.

In the future, however, much bigger volumes can be surveyed with the
GAIA satellite
(http://astro.estec.esa.nl/ GAIA/Science/), which will have improvements of
factors of 100 in accuracy and 1000 in limiting magnitude over
Hipparcos (http://astro.estec.esa.nl/GAIA/Science/Science.html).
Anomalous associations of solar-type stars will then be detectable out
to 1.7 kpc since the sensitivity of the GAIA parallax measurements will be
good enough to determine the distances of stars this far away given their
apparent magnitudes.   
Such stars ($B<17$) will easily be bright enough for precise velocity
dispersion measurements.  

\subsection{Summary}

If dark galaxies contain only, CDM then strong gravitational lensing 
(Section 3.1.1) is required for a direct detection 
unless there is a fortuitous discovery of an anomalous stellar
association (Section 3.5).  If the dark galaxies have NFW density profiles,
then strong lensing measurements will, however, be extremely difficult.
More likely, indirect methods  
(Sections 3.1.2, 3.1.3, 3.3 or 3.4) are required.
If some small fraction of the dark matter is baryonic, then far-infrared
(Section 3.2.1) or near-infrared (Section 3.2.2) searches may be productive
if the dark matter is brown dwarfs or CWDs, respectively, although in the
latter case the CWDs would need to be fairly concentrated near the galaxy
centers.

\section{Where to look for dark galaxies}

For the direct (gravitational lens and infrared
search) methods outlined in previous section, 
it is clear that detecting dark galaxies with present technology is
marginal at best.  This means that any such observations are best performed 
by concentrating on a small number specific target fields 
for which there is a high probability of a dark galaxy
existing.  We suggest the following possibilities.

\subsection{Blue compact dwarf and isolated HII galaxy fields}

Blue compact dwarfs (BCDs) and low-luminosity HII starburst galaxies are small
galaxies which are undergoing a sudden burst of star formation.  They
often have very disturbed morphology; a prototype is UGC 6456
(see Lynds et al.~1998 for an optical $HST$ image).  They typically have
star formation rates so high that they could not have maintained them for
any considerable time or else they would have a total mass in stars higher
than is permitted given their total optical 
luminosity.  

Something must have triggered these objects into bursting.  
Both HI (Taylor et al.~1995) and optical (Telles \& Maddox 2000)
analyses show that 
although a handful of HII galaxies have companions,
most appear to be quite isolated.  
The analysis of Telles \& Maddox (2000) in particular
probes down to very faint limits.  Their results suggest that something
other than an interaction with a normal luminous galaxy is responsible for
triggering these isolated HII galaxies into bursting.  One
possibility is an interaction with a dark galaxy that neither they
nor Taylor et al.~see because it has too little or no stars or gas.    
Such a dark galaxy will have gravitational effects that mimic the effects of
the companion in the few cases where the HII galaxies {\it does} have one. 
Therefore fields around isolated BCDs and HII galaxies would appear to be
good target fields. 

\subsection{Cores of poor clusters of galaxies}

An environment where many dark galaxies probably exist is in the cores of
poor, dynamically unevolved clusters.  This is because the small halos there
have not been able to form stars, for the reasons given in Section 1, yet
the overall density (as inferred by the luminous galaxy density
and/or dynamical considerations) is high 
enough that the number density of small halos will not be extremely small.
The Ursa Major Cluster is a good example of such an
environment, particularly since deep optical observations (Trentham et
al.~2000) show that dwarf galaxies of the type that are so numerous in
Virgo (Phillipps et al.~1998) do 
not exist here, despite the cluster being only
one-twentieth as massive as Virgo (Tully et al.~1996).

These ideas can be combined with those in the previous section.  We therefore
suggest that some of the best fields of all to target are fields in the
vicinity of starbursting dwarf galaxies in the centers of poor clusters.
The field around the BCD galaxy umd 38 (Trentham et al.~2000)
in the center of the Ursa Major Cluster is a particularly good example. 

\subsection{Outflows from isolated galaxies}

\begin{figure}
\begin{center}
\vskip1mm
\epsfig{file=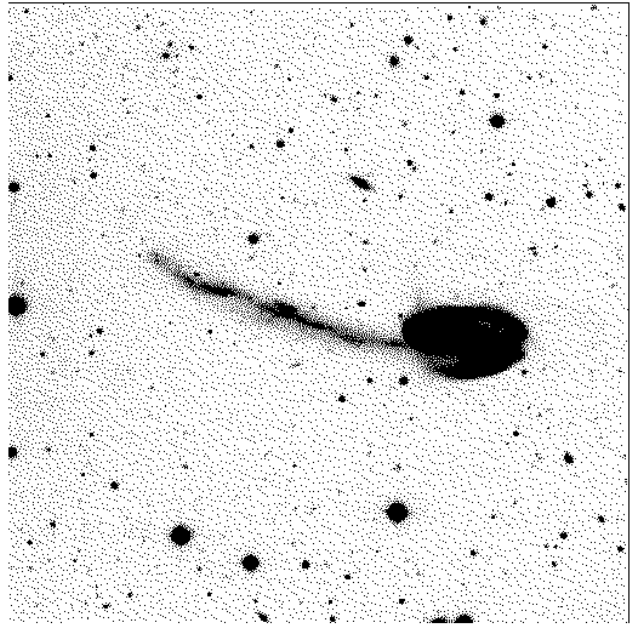, width=8.4cm}
\end{center}
\vskip-3mm
\caption{
A $g$-band image of UGC 10214 from the INT Wide Field Survey
(http://www.ast.cam.ac.uk/$^{\sim}$wfcsur).
North is up and east is to the left.
The image is 6.2 arcminutes $\times$ 6.2 arcminutes.
The galaxy is at a redshift of $z=0.03$ (Bottinelli et al.~1993),
at which 1 arcminute corresponds
to 40 kpc.
}
\end{figure}

Interacting galaxy systems are common (see e.g.~{\it The
Atlas of Peculiar Galaxies}; Arp 1966), and often show evidence
for material flowing between galaxies.
A rarer phenomenon is material flowing out of a galaxy towards
apparently nothing.  The outflow from UGC 10214 (see Figure 4) is a
good example.  It is possible that material is being gravitationally
pulled out of the galaxy by a dark companion, just like material is pulled
out of one member in a more normal interacting system by another
member.  We therefore suggest that the
areas of the sky at the far end of the streams from the host galaxies
in systems like this would 
also be good target fields. 

\subsection{Fields of dwarf galaxies with anomalous velocities}

Some small galaxies have velocities that are difficult to
explain given their location in space.  An example is Leo I, which is
close to the Milky Way Galaxy (MWG)
but has an anomalously large velocity with
respect to the MWG (Zaritsky,
et al.~1989, Byrd et al.~1994).  It is
probably different from the other Galactic satellites in that it is
not orbiting the MWG.

One possibility is that Leo I {\it was} orbiting the MWG at some time in the
past but recently received a velocity kick from a dark companion that we
do not see.  If we could identify where in the sky such a companion might
be, this might be a reasonable target field for an infrared search for
dark galaxies of the type described in the previous section.  The region
of the sky where such a companion could reside is not large since tidal
torques are a rapidly-falling ($\sim 
r^{-3}$) function of distance; a
hypothetical dark companion must therefore be very close to Leo I.
Leo I is just one example.  As optical spectroscopic surveys get more
sensitive, perhaps similar objects (low-mass galaxies are best
since they experience greater accelerations for torques of a given
magnitude)
could be found at greater distances 
where gravitational lensing searches could also be used
(a dark companion to Leo I would be too close even if
its dark matter profile is very
steeply peaked; see Section 3.1.1). 

While we suggest this as a possible target field for a 
dark galaxy, this interpretation of the unusual velocity of Leo I is
far from secure.  
An alternative possibility is that Leo I is just passing through the
Local Group and was never associated with the MW.  This possibility 
might be suggested by the fact that it may have no old stars at all
(Lee et al.~1993, Hodge 1994) unlike other dwarf elliptical 
satellites of the MW such as Ursa Minor (Olszewski \& Aaronson 1985).  

\subsection{Fields of intergalactic planetary nebulae and/or red giants}

Intergalactic red giant stars (Ferguson, Tanvir \& von Hippel 1998)
and planetary nebulae (Theuns \& Warren 1997; Freeman et al.~2000) are
known to exist in the Virgo and Fornax galaxy clusters
and are thought to have been released from their parent
galaxies at some time in the past.

Such objects in the true field are not known but if discovered, the
areas around these objects would present
reasonable target fields for a dark galaxy search.  The idea would be that
these objects might have been accreted by a late-forming dark halo as
described in Section 3.5.  Certainly if two or more intergalactic stars 
of any type that are close in physical space can be identified, this would
give an excellent target field.

There are a number of theoretical and observational hints 
that large numbers of low-mass
galaxies composed entirely of dark
matter exist in the field.  The theoretical considerations follow
from the prediction of cold dark matter theory that there exist
many low-mass galaxies for every massive one. 
The observational considerations follow 
from the observed paucity of these 
low-mass galaxies in the field but not in dense clusters of galaxies; this
suggests that the lack of small galaxies in the field is due to the
inhibition of star formation in the galaxies as opposed to the fact
that their small dark matter halos do not exist.  In this work we 
outline the likely properties of low-mass dark galaxies, and
describe observational strategies for finding them, and where in
the sky to search.  The results are presented as a function of the global
properties of dark matter, in particular the presence or absence of a
substantial baryonic dark matter component.  If the dark matter is
purely cold and has a Navarro, Frenk \& White density profile, directly
detecting dark galaxies will only be feasible with
present technology if 
the galaxy has a maximum velocity dispersion in excess of 70
km s$^{-1}$, in which case the dark galaxies
could strongly lens background objects.  This is much higher than the
maximum
velocity dispersions in most dwarf galaxies.  If the dark matter in galaxy
halos has a baryonic component close to the cosmic ratio, the possibility
of 
directly detecting dark galaxies is much more realistic; the optimal method of
detection will depend on the nature of the dark matter. 
A number of more indirect methods are also discussed.
\section*{Acknowledgments} 

Helpful discussions with Simon Hodgkin throughout this project are
gratefully acknowledged.  
This research has made use of the NASA/IPAC Extragalactic Database (NED),
which is operated by the Jet Propulsion Laboratory, Caltech, under agreement
with the National Aeronautics and Space Association.  
and of the INT Wide Field Survey
(http://www.ast.cam.ac.uk/$^{\sim}$wfcsur).

\end{document}